\documentclass[pre,amsfonts, amssymb, amsmath, reprint]{revtex4-2}
\usepackage[english]{babel}
\usepackage[utf8]{inputenc}
\usepackage[normalem]{ulem}
\usepackage{amsthm}
\usepackage{mathtools}
\usepackage{physics}
\usepackage{xcolor}
\usepackage{graphicx}
\usepackage{placeins}
\usepackage{csquotes}
\usepackage{soul}
\usepackage{hyperref} 
\usepackage[colorinlistoftodos, color=green!40, prependcaption,textsize=footnotesize]{todonotes}

\begin{document}
\title{Modulation of ionic conduction using polarizable surfaces}
\author{A. P. dos Santos$^{1,2}$, F. Jim\'enez-\'Angeles$^{2}$, A. Ehlen$^{3}$, and M. Olvera de la Cruz$^{2,3,4}$}
\affiliation{$^{1}$Instituto de F\'isica, Universidade Federal do Rio Grande do Sul, Caixa Postal 15051, CEP 91501-970, Porto Alegre, RS, Brazil.}
\affiliation{$^{2}$Department of Materials Science and Engineering, Northwestern University, Evanston, Illinois 60208, USA}
\affiliation{$^{3}$Applied Physics Program, Northwestern University, Evanston, Illinois 60208, USA}  
\affiliation{$^{4}$Department of Physics and Astronomy, Northwestern University, Evanston, Illinois 60208, USA}
 



\date{\today} 

\begin{abstract}
Hybrid ionic-electronic conductors have the potential to generate memory effects and neuronal behavior. The functionality of these mixed materials depends on ion motion through thin polarizable channels. Here, we explore different polarization models to show that the current and conductivity of electrolytes is higher when confined by conductors than by dielectrics. We find non-linear currents in both dielectrics and conductors, and we recover the known linear (Ohmic) result only in the two-dimensional limit between conductors. We show that the polarization charge location impacts electrolyte structure and transport properties. This work suggests a mechanism to induce memristor hysteresis loops using conductor-dielectric switchable materials.
\end{abstract}


\maketitle
\section{Introduction} 

Switchable conductance is desirable for designing densely interconnected (neuromorphic) systems for information storage, performing complex logic operations, and executing neural network algorithms ~\cite{yang2013memristive,van2018organic,science.1254642}. To emulate neural activity, such as voltage spiking~\cite{RoKa21,PhysRevLett.130.268401} and synaptic plasticity~\cite{Kuzum_2013,RoEm23}, researchers use
different materials considering  switchable ionic or electronic conduction~\cite{adfm.201100686,sangwan_multi-terminal_2018}. For example, the gate resistance tunability has been explored for neuromorphic circuits using monolayer MoS$_2$ multi-terminal memtransistors~\cite{sangwan_multi-terminal_2018}. To expand functionality and flexibility of device design,
integration of ionic and electronic conduction is an attractive option as it may allow for imitating synaptic potentiation, emulating plasticity  \cite{aelm.202100866,chemrev.1c00597}, and achieving neural interfacing. The coupling between ionic and electronic transport is promising for leveraging other applications such as biosensing, energy storage, and responsive materials.

Mixed ionic-electronic conductors are materials that conduct both ions and electronic charge carriers (electrons and/or holes)\cite{RIESS20031}. Recent developments combine electronic and ionic conductor materials into alternating layers of nanometric dimensions where the ionic and electronic conduction occur simultaneously~\cite{rivnay2016structural}.
The close proximity of ionic and electronic charge carriers in these devices means that the conduction behavior of one influences the other. These materials' electronic properties can be modified by the stoichiometry~\cite{kim_influence_2014,qiu_hopping_2013}, gate biasing~\cite{sangwan_multi-terminal_2018, Mischa23}, and structural changes~\cite{rivnay2016structural}.
The coupled ionic and electronic interactions and transport need special understanding beyond the comprehension of their independent behavior. 

Despite numerous technological applications envisioned by using mixed ionic-electronic conduction, the lack of fundamental understanding impedes rational materials design. 
One essential component of mixed conduction is the effect of induced electronic polarization on ionic conduction.
Previous work showed that the electronic properties of surfaces affect the nearby ions, specifically via the induced polarization charges due to dielectric mismatch on the material-electrolyte interface~\cite{LiMe13,schlaich2022electronic,KaRo22,jimenez2023faraday}. A recent formalism employs the Thomas-Fermi model~\cite{PhysRevLett.31.681} to consider polarization effects on the ions' transport in strong confinement by dielectrics and conductors~\cite{KaRo22}. However, the formalism is derived only for ions constrained to move in two dimensions. Using two different approaches, here we study ion conduction in strong confinement in a slit-like channel and we recover some aspects of the two-dimensional behavior predicted by the Thomas-Fermi model.

Frameworks and models that integrate the interfacial electronic polarization of materials and electrostatic molecular interactions are essential for exploiting the properties of interfaces in nanometer slit confinement~\cite{son2021image,schlaich2022electronic,jimenez2023faraday}. 
As illustrated in Fig.~\ref{fig_app}(a), the electrostatic potential generated by a charge in the center of a narrow slit-like channel is highly dependent on the polarizability of the channel. Whereas the range of the potential is increased by dielectric confinement, the potential in strong conducting confinement is so screened that it can be considered short range. Here,  we study ionic conduction in strongly confining slit-like channels and show that the ionic conductivity and the ionic adsorption are significantly impacted by changes in the confining material polarization (from dielectric and conductor), the confining distance, and the location of the polarization charges. Results from density functional theory show that the polarization plane location is a material- dependent property~\cite{lang_theory_1973}. Hence, by adjusting the polarization plane location we consider materials of different electronic properties.


The paper is organized as follows. First, we develop an efficient method for treating ionic interactions in strong confinement by conductors and define the system parameters. We then find a nonlinear ionic conduction response, which is a prerequisite for neuromorphic behavior. Finally, we demonstrate that the location of the polarization charge impacts the ionic distribution and transport through the channel. The placement of the polarization charge offers a mechanism for modeling material-dependent properties of polarizable surfaces.

\section{Green's Function Methods and System Parameters}

\begin{figure}[t]
\begin{center}
\includegraphics[scale=0.32]{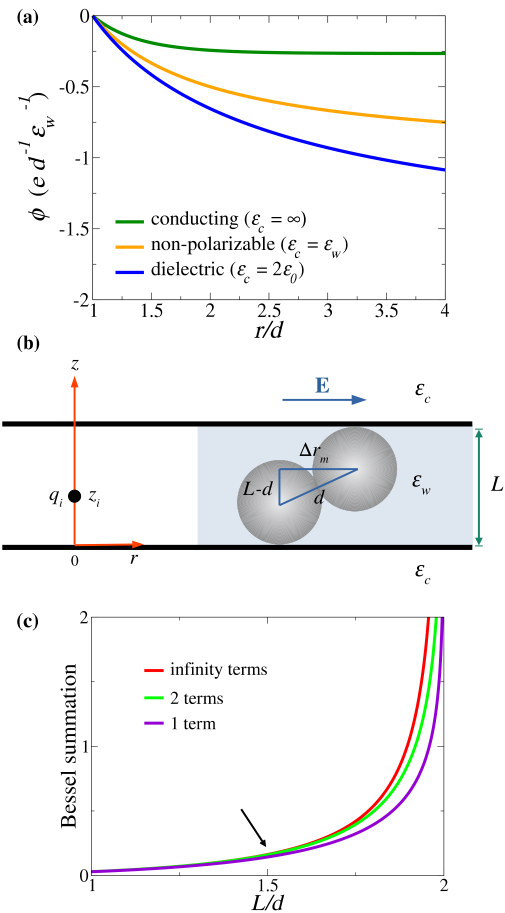}
\end{center}
\caption{Ions strongly confined by polarizable surfaces. (a) Electrostatic potential in the radial direction ($r$) for $\varepsilon_w = 5 \varepsilon_0$, from an ion at the center ($z=L/2$, where $L=1.475d$) of two conducting, non-polarizable, or dielectric surfaces. We use a $1/r$ potential for non-polarizable confinement and the traditional method of images for the dielectric and conducting confinements. To aid visualization, the potential is set to zero at $r/d=1$. (b) The cylindrical coordinate system used to derive the FSCC model, and the setup employed in molecular simulations; $z$ and $r$ are the axial and radial coordinates, respectively, and $\Delta r_m$ is the minimal ion-ion radial separation; {\bf E} is the external electric field applied to investigate the ions' transport in the direction parallel to the surfaces. (c) Convergence of the Bessel summation $\sum_{n=1}^{\infty}K_0(\frac{n\pi\Delta r_m}{L})$ in Eq.~\eqref{eq:SCMphi} for various interplane distances $L$ at the minimum ion-ion separation $\Delta r_m=\sqrt{L(2d-L)}$. The black arrow indicates the superior limit we consider the function is converged using just one term in sum.}
\label{fig_app}
\end{figure}

We consider a system consisting of a 1:1 electrolyte with $N_+$ cations and $N_-$ anions of diameter $d$ confined between two polarizable surfaces placed parallel to  the $x$-$y$ plane and separated by a distance $L$ in the $z$-direction. The (periodic) box has side lengths $L_x$ and $L_y$ in the $x$ and $y$ directions. Each region has a uniform dielectric constant, which is $\varepsilon_{\rm w}$ and $\varepsilon_c$ for the electrolyte and confining material, respectively. To investigate the ions' transport, we apply an external field {\bf E} in the direction parallel to the confining walls. The setup is shown in Fig.~\ref{fig_app}(b). 

The ion-ion electrostatic interaction that accounts for the polarization of the confining conducting surfaces is derived from the method developed in Ref.~\cite{GiDo17}. It circumvents explicit calculation of induced surface charge, applies to strong confinement ($1<L/d < 2$), and consists of a single term for $1<L/d \lesssim 1.5$. The method takes advantage of the short-ranged ion-ion electrostatic interaction in strong confinement by conductors, which allows us to use the minimum image convention rather than Ewald summation methods to compute the electrostatic interactions. This significantly accelerates molecular simulations, so we refer to the method as the \textit{fast strong conducting confinement} (FSCC) method. To derive it, we consider a single confined charge $q_i$ at ${\bf r}_i=(0,z_i)$, on the $z$ axis of the cylindrical coordinate system, see Fig.~\ref{fig_app}(b).
The polarizable infinite planar surfaces are placed at $z=0$ and $z=L$. The Poisson equation was solved for this setup in the context of confined ionic liquids~\cite{GiDo17}.
The electrostatic potential at an arbitrary position ${\bf r}$, generated by the ion $q_i$ is given by
\begin{equation}
\label{eq:SCMphi}
\phi({\bf r})=\frac{4 q_i}{\varepsilon_{\rm w} L}\sum_{n=1}^{\infty}\sin{\left(\frac{n \pi z}{L}\right)}\sin{\left(\frac{n \pi z_i}{L}\right)}K_0\left(\frac{n\pi\Delta r}{L}\right) \ ,
\end{equation}
where $K_0$ is the modified Bessel function of order $0$ and $\Delta r=\sqrt{(x-x_i)^2+(y-y_i)^2}$.
For strong confinement, meaning $L$ of order of $d$, just the first term ($n=1$) of the summation in Eq.~\eqref{eq:SCMphi} is necessary because the minimal separation between two ions in the radial direction is $\Delta r_m=\sqrt{L(2d-L)}$, see Fig.~\ref{fig_app}(b). This condition implies that the argument of the modified Bessel function is at least $n\pi \sqrt{(2d/L-1)}$, leading to a fast convergence of sum. Considering $L=1.475 d$, this gives $K_0(n\pi \sqrt{0.356})\approx 0.13, 0.01, 0.001$, for $n=1,2,3$, respectively. For the systems studied in this work, we use just the first term, $n=1$, leading to simple expressions, derived below. For larger $L$, $1.5<L/d<2$, one needs to consider more terms in the summation, see Fig.~\ref{fig_app}(c).

The force between two ions $q_i$ and $q_j$ at positions $\vec{r}_i$ and $\vec{r}_j$, converting to Cartesian coordinates, is:
\begin{equation*}
F_x^{(i,j)}=\frac{4 \pi q_i q_j (x_i-x_j)}{\varepsilon_{\rm w} L^2 \Delta r}\sin{\left(\pi \frac{z_i}{L}\right)}\sin{\left(\pi \frac{z_j}{L}\right)}K_1\left(\pi\frac{\Delta r}{L}\right) \ ,
\end{equation*}
\begin{equation*}
F_y^{(i,j)}=\frac{4 \pi q_i q_j (y_i-y_j)}{\varepsilon_{\rm w} L^2 \Delta r}\sin{\left(\pi \frac{z_i}{L}\right)}\sin{\left(\pi \frac{z_j}{L}\right)}K_1\left(\pi\frac{\Delta r}{L}\right) \ ,
\end{equation*}
\begin{equation*}
F_z^{(i,j)}=-\frac{4 \pi q_i q_j}{\varepsilon_{\rm w} L^2}\cos{\left(\pi \frac{z_i}{L}\right)}\sin{\left(\pi \frac{z_j}{L}\right)}K_0\left(\pi\frac{\Delta r}{L}\right) \ ,
\end{equation*}
where $F_x^{(i,j)}$, $F_y^{(i,j)}$ and $F_z^{(i,j)}$ are the $x$, $y$, and $z$ components of electrostatic force and $K_1$ is the modified Bessel function of order $1$.

The self-electrostatic interaction describes the interaction between an ion and the surface charge it induces. It would be computed by Eq.~\eqref{eq:SCMphi} as $r \rightarrow 0$, but $K_0$ diverges in this limit. In this case, we consider the integral based method~\cite{GiDo17}, which gives
\begin{equation*}
\phi_{self}(z_i)=\frac{q_i}{\varepsilon_{\rm w}}\int_0^\infty  \frac{2 e^{-2 k L}-e^{-2 k z_i}-e^{2 k z_i-2 k L}}{(1-e^{-2 k L})} dk \ .
\end{equation*}
The self-force acting on charge $q_i$ in the $z$ direction is $F^{self}_z=-\frac{q_i}{2}\frac{\partial}{\partial z_i}\phi_{self}(z_i)$, which can be written
\begin{eqnarray}
F^{self}_z=\frac{q_i^2}{4 L^2 \varepsilon_{\rm w}}\bigg[ \psi^{(1)}(1-z_i/L)-\psi^{(1)}(z_i/L) \bigg] \ ,
\end{eqnarray}
where $\psi^{(1)}$ is the polygamma function of first order.

To study ionic transport in confinement by conductors, we incorporate the FSCC method into molecular dynamics simulations. Additionally, we consider the case of confining dielectric walls using the method of periodic Green function (PGF)~\cite{DoGi17}. While both Green function methods (FSCC and PGF) agree for conducting surfaces, the FSCC method is around two orders of magnitude faster. 

We investigate a $1:1$ electrolyte under two confinement widths, $L = 1.1d$, confining the ions almost exactly to a plane, and a larger value $L=1.475d$. The dielectric constant within the slit is $\varepsilon_{\rm w} = 5\varepsilon_0$ to represent an organic solvent or strongly confined water~\cite{Fumagalli2018}. We set $N_+=N_-=10$, and each ion has a charge $q_+=e$ or $q_-=-e$ located at its center, where $e$ is the positive elementary charge. The lateral dimensions of the simulation box are $L_x = 11.9 d$ and $Ly=12d$, which were chosen based on surface discretization that will be relevant in Section IV. The ions' excluded volume is represented with a truncated Lennard-Jones potential (also known as the Weeks-Chandler-Anderson potential) using an energy scale $\epsilon_{\rm LJ}$, with $\sigma_{\rm LJ-ion} = d$. The walls confine the ions via a truncated Lennard-Jones potential, with $\sigma_{\rm LJ} = 0.8353d$ and $\epsilon_{\rm LJ}$. The wall-ion interaction is calculated using the Lorentz-Berthelot mixing rule. The centers of the confining walls are located at $z=-0.5 \sigma_{\rm LJ}$ and $z=L+0.5 \sigma_{\rm LJ}$. The values for the energy, mass, and distance scales are $\epsilon_{\rm LJ} = k_B T$ where $T = 298$ K, the ion mass set to the mass of sodium $m = 22.98$ amu, and $d = 0.425$ nm, a typical size of a hydrated ion. We assume symmetry of ion mass and diameter to focus on the effect of polarization charge on ion transport. Effects due to asymmetry of other ion properties will be explored in future work. We study the ionic currents by applying an electric field $\mathbf{E}=E \hat{x}$ tangentially to the surfaces, see Fig.~\ref{fig_app}(b), and compare the results for conducting ($\varepsilon_c\rightarrow\infty$), dielectric ($\varepsilon_c = 2\varepsilon_0$), and non-polarizable surfaces.

We perform molecular dynamics simulations using the Langevin method with a damping parameter of 100 fs and periodic boundary conditions in $x$ and $y$ directions.
We consider $100000$ MD steps for equilibration and $100$ MD steps of space between $10000$ uncorrelated samples created for further analysis.



\section{Effects of confinement and material polarizability}

\begin{figure}[t]
\begin{center}
\includegraphics[scale=0.29,trim={0.7cm 0.9cm 0.8cm 0.8cm},clip]{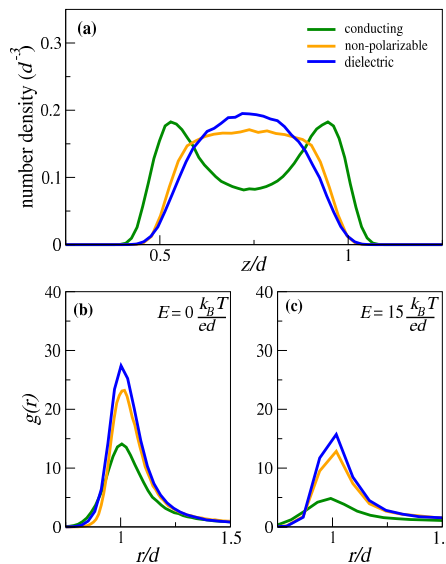}
\end{center}
\caption{Results for $1:1$ electrolyte for various surface polarization conditions for $L=1.475 d$. (a) The ionic concentration profiles. (b) Cation-anion radial distributions with no applied electric field. (c) Cation-anion radial distributions  with applied electric field $E=15~k_BT e^{-1}d^{-1}$.}
\label{fig_prof2A}
\end{figure}

\begin{figure}[t]
\begin{center}
\includegraphics[scale=0.32,trim={0.6cm 0.9cm 0.8cm 0.2cm},clip]{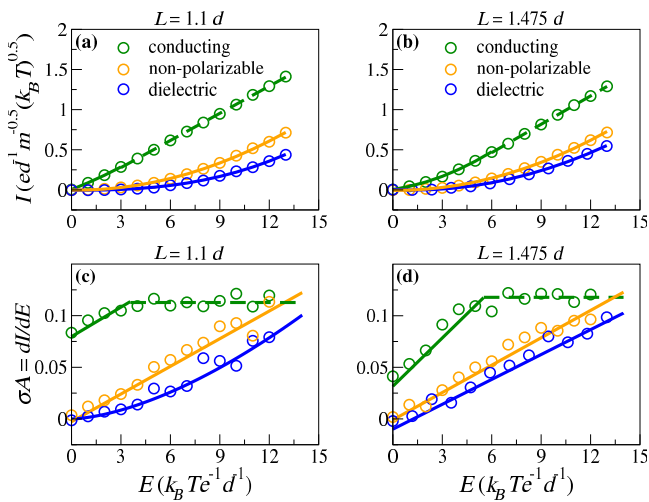}
\end{center}
\caption{Results for $1:1$ electrolyte for various surface polarization conditions. Current as a function of applied electric field for separation (a) $L=1.1 d$ and (b) $L=1.475 d$. The electrolyte conductivity $\sigma A$ as a function of applied electric field for separation (c) $L=1.1 d$ and (d) $L=1.475 d$. The two regimes for conducting cases can be observed for each curve with dashed and full lines, which are fits of current curves. The current is calculated as $I=\langle\sum_{i=1}^{N_++N_-}q_iv_{ix}/L_x\rangle$, where $q_i$ and $v_{i}$ are the charge and the velocity of particle $i$.}
\label{fig_prof2B}
\end{figure}

We analyze the ionic density profiles, the current $I$, the radial distributions, and the conductivity of the confined ions. Fig.~\ref{fig_prof2A}(a) shows that the ions are more adsorbed to conducting surfaces and more repelled from dielectric surfaces than they are from non-polarizable surfaces. This behavior is expected since the ionic interactions with dielectric or conducting surfaces can be understood in terms of equally or oppositely charged images, respectively~\cite{jackson1998}. 
Fig.~\ref{fig_prof2A}(b) shows the pair correlation functions between oppositely charged ions, $g(r)$, which is related to the potential of mean force between ions $w(r)$ as $g(r)=e^{-w(r)/k_BT}$. 
Therefore, Fig.~\ref{fig_prof2A}(b) shows that the effective interaction between oppositely charged ions is modified by the properties of the confining walls. The ion pairing is seen as a peak in the anion-cation radial distribution function at $r/d \approx 1$. The peak  decreases by a factor of $\approx 2. 5$ for conducting surfaces compared to dielectric surfaces (see Fig.~\ref{fig_prof2A}(b)), signifying that the formation of pairs is less favorable between conducting surfaces than dielectrics. The decrease in pair formation in conducting confinement occurs due to the difference in ion-ion interactions near dielectric and conducting surfaces. An ion $q_i$ interacts with an oppositely charged ion $q_j$ {\it and} with $q_j$'s equally charged image charges near a dielectric material, {\it but} mostly with the $q_j$'s opposite image charges near a conducting material.
An external electric field {\bf E} (applied in the channel's surface parallel direction) decreases the effective anion-cation attraction, but the effect due to the confining walls' polarization persists (see Fig.~\ref{fig_prof2A}(c)).


Fig.~\ref{fig_prof2B}(a) and (b) show that the currents obtained with confining conductive surfaces are much higher than those obtained with dielectric surfaces.
The applied electric field and the current are related as $I=A \sigma E$, where $\sigma$ is the conductivity and $A$ is the slit's cross-sectional area. A field-dependent conductivity (non-linear $I$-$E$ relationship) is essential in {\em memristors}~\cite{RoEm23}, which are the basis of memory systems.
The three $I$-$E$ curves and their corresponding conductivities in Fig.~\ref{fig_prof2B} exhibit diverse regimes, depending on separation between surfaces and polarization. In the field range studied here, for a given value of $L$, the ionic current and conductivity are higher between conductors than between dielectric or non-polarizable surfaces.  At the separation distance of $L=1.1d$ between conductors, the current is linear in almost the entire range of studied electric fields. Under such strong confinement, ions are constrained to move nearly on a plane, so the Ohmic behavior in our simulations is consistent with the two-dimensional model between conducting surfaces~\cite{KaRo22}. The ionic current between conductors at the larger separation distance ($L=1.475d$) shows a non-linear trend for fields below $E \sim 6$ $k_BTe^{-1}d^{-1}$.
For non-polarizable and dielectric surfaces, the $I$-$E$ curves present a quadratic form, which reflects the linear curves for the conductivity. The exception occurs for dielectric confinement at short separations, which data are best fitted with a $E^{2.6}$ function, which gives a $E^{1.6}$ dependence for conductivity.
Combining results using conductor to non-conductor switchable materials suggests a  mechanism for inducing hysteresis loops~\cite{RoEm23} in memristors.

The current nonlinear behavior can be understood in terms of the Onsager’s ion pairing theory \cite{Onsager1934,RoKa21}. According to this theory, the ions form pairs that last for a time $\tau_d$ and stay free for a time $\tau_a$.  Ion pairs have zero net charge, meaning they do not contribute to the current. Hence, a larger number of ionic pairs between dielectrics than between conductors decreases the overall current. With the increase in electric field the ion pair's duration time $\tau_d$ decreases (reflected as a decrease of the  radial distribution peak in Figs.~\ref{fig_prof2A}(b,c)) and the current increases. Due to the weaker attraction in confinement by conductive surfaces, $\tau_d$ is shorter  than between dielectric and non-polarizable surfaces. Therefore, in confinement by conductors the ions tend to move dissociated as an electron gas (Ohmic behavior) whereas in confinement by dielectric and non-polarizable surfaces the ion pairs persist in a broader range of electric fields. At high enough $E$-fields all confining materials give an Ohmic response. This variety of behaviors show that the particle-particle interaction strongly influences the  ionic current revealing the modulation of surface properties to be a tool for tuning ionic behavior.



\begin{figure}[t]
\begin{center}
\includegraphics[scale=0.35,trim={0.35cm 0.2cm 0.1cm 0},clip]{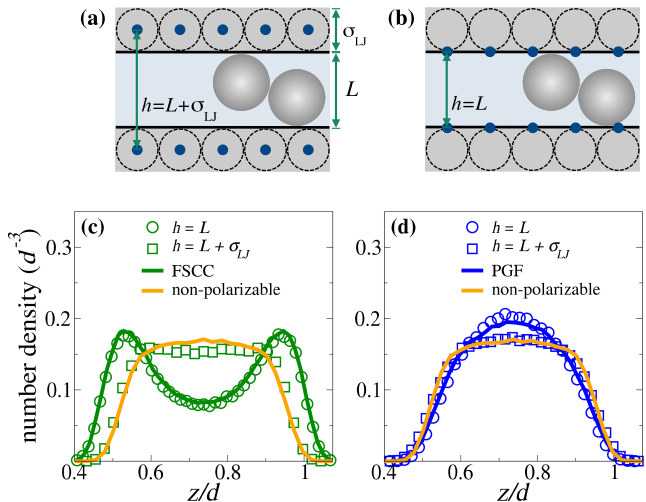}
\end{center}
\caption{
Two models for charge placement, using different charge separation distances $h$: (a) $0.5 \sigma_{\rm LJ}$ {\em inside} the surfaces ($h=L+\sigma_{\rm LJ}$) and (b) exactly on the {\em surface} of the material ($h=L$); Ion number density profiles for both charge placement models, compared with exact polarizable and non-polarizable results for (c) conducting and (d) dielectric surfaces. In both figures, \textit{inside} charge placement causes the ions to behave almost as if the surfaces were non-polarizable.
\label{fig_comp}}
\end{figure}
%

\section{Effects of the polarization plane placement}

The Green function methods apply to planar confinement. For other geometries, such as conical channels suggested for current rectification~\cite{GaKo18,boon_pressure-sensitive_2022}, other methods are necessary.
\textit{Explicit polarization} methods incorporate the surface polarizability into ion-ion interactions by calculating the induced charge in a way that is not restricted to a specific geometry~\cite{electrode2022,nguyen2019incorporating}.
Conductors are implemented by imposing a uniform surface potential  $\psi({\bf s})=const$.  In dielectrics with no free surface charges, the electric field boundary conditions at the interface between two media are given by $\varepsilon_{\rm w} \mathbf{E}_{\rm w,n}=\varepsilon_{\rm c} \mathbf{E}_{\rm c,n}$ and $\mathbf{E}_{\rm w,t}= \mathbf{E}_{\rm c,t}$, where the subscripts $w$ and $c$ refer to the ions' solvent and confining material, respectively, and $n$ and $t$ indicate the perpendicular and tangential electric field components.

The electrostatic boundary conditions define the polarization plane (image plane) which represents the location of the polarization charges. 
In early density functional theory studies, Lang and Kohn found that the image plane at metallic surfaces is located at the centroid of the induced charge profile \cite{lang_theory_1973}. Later work shows that the induced polarization charge peak resides in a range from tenths of an angstrom inside the surface nuclei to a few angstroms outside, depending on electron density of the material, external electric field, and other factors \cite{luque_electric_2012}. More recently, it has been suggested that the image plane location
affects the adsorption of charged peptide species on gold \cite{heinz_polarization_2010} and shown that the double layer capacitance of silver and graphite surfaces depends on image plane placement \cite{schmickler_electronic_2020}. 

In classical molecular simulations, however, the induced polarization charge is frequently placed at the center of the beads forming the surface \cite{jimenez2023faraday,nguyen2019incorporating}. For conducting surfaces, and extending the concept to dielectrics, we study the impact of this choice by placing  the polarization charges (1) \textit{inside} the channel walls ($h=L+\sigma_{\rm LJ}$, Fig.~\ref{fig_comp}(a)) or (2) on the \textit{surface} ($h=L$, Fig.~\ref{fig_comp}(b)). Because our methods require explicit surface charge to be represented on a discrete mesh, we locate the polarization charges on a hexagonal graphene structure with an average nearest-neighbor distance of $0.332d$  on the $x$-$y$ plane. All the other parameters are the same as described in Section II. We perform molecular dynamics simulations for only the case of $L = 1.475d$ using the polarization methods implemented in  LAMMPS~\cite{thompson_lammps_2022}. In conductors, a constant surface potential is maintained using the Gaussian charge model  \cite{siepmann1995influence,reed2007,electrode2022}. For dielectrics, the electrostatic boundary conditions are considered using a boundary element method~\cite{nguyen2019incorporating}. Further details of the wall and conductor models are supplied in the Supplemental Material~\cite{sm}.

\begin{figure}[t!]
\includegraphics[scale=0.315,trim={0.9cm 0.35cm 0.4cm 0.55cm},clip]{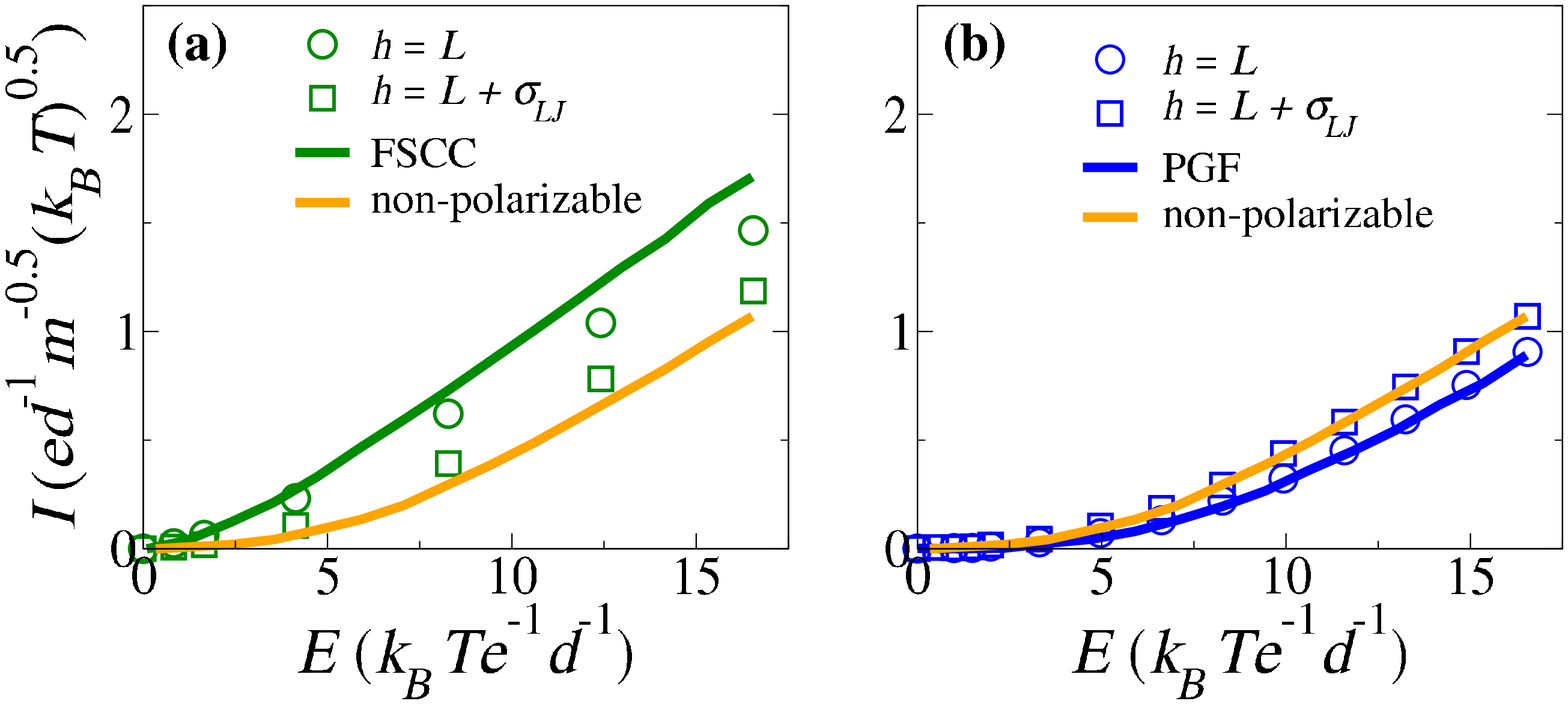}
\includegraphics[scale=0.306,trim={0.1cm 0.2cm 0.1cm 0.2cm},clip]{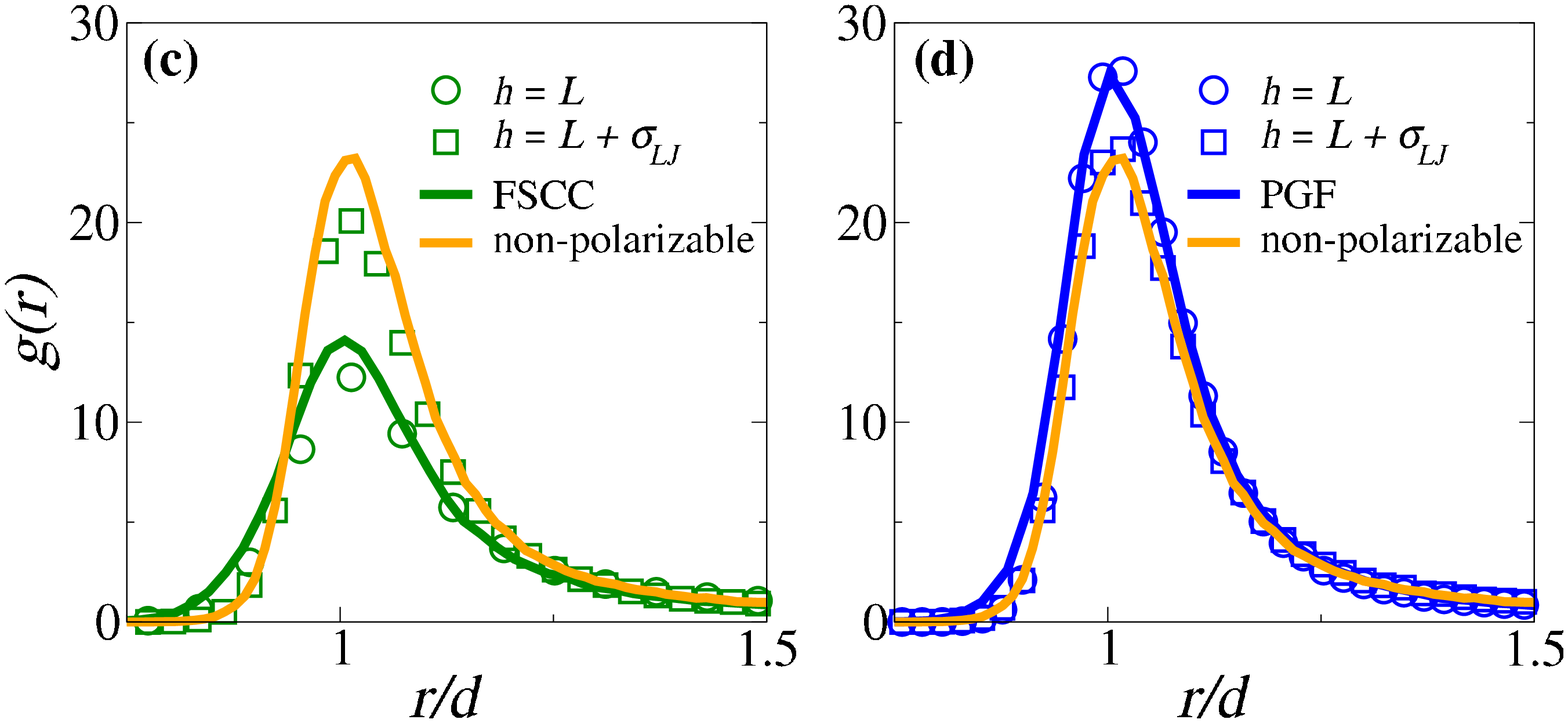}
\caption{Ionic current for confining (a) conducting and (b) dielectric surfaces. Cation-anion radial distribution for (c) conducting and (d) dielectric surfaces, with no applied electric field. Analogously to Fig.~\ref{fig_comp}, ion current is much more significantly impacted by changing surface material for \textit{surface} charge placement.}
\label{fig_comp2}
\end{figure}

Fig.~\ref{fig_comp}(c) and (d) show
the ion density profiles for conductive and dielectric surfaces, respectively. The figures include the profiles at the polarization charge locations of $h=L$ and $h=L+\sigma_{\rm LJ}$. For comparison, we include the results from the Green function methods (FSCC and PGF) and for non-polarizable systems. Placing the polarization charges on the walls' \textit{surface} ($h=L$) leads to conductive surfaces adsorbing the ions and dielectric surfaces repelling them.
For both dielectrics and conductors, the effect of surface polarization is significantly dampened when the polarization charges are located inside the surface. Placing the polarization charges \textit{inside} the surfaces causes the profiles of polarizable systems to converge towards those of non-polarizable systems, while placing the polarization charges on the \textit{surface} results in profiles that overlap with the Green function methods. 
Similarly, the ionic current in polarizable models is close to that of non-polarizable system (see Fig.~\ref{fig_comp2}(a) and (b)), when the polarization plane is \textit{inside} the surfaces. When the polarization plane is placed on the walls' \textit{surface}s, the current is similar to that from the Green function methods. We attribute the slight discrepancy between the two methods at high fields to the discrete mesh employed in the \textit{explicit polarization} models (see Fig.~S2 in the Supplemental Material~\cite{sm}). The pair correlation functions in Figs.~\ref{fig_comp2}(c) and (d) explain the current behavior. The systems with \textit{surface} charge placement indicate decreased ion clustering (lower peak) for conductive surfaces and enhanced ion clustering (higher peak) for dielectric materials, aligning with the Green function methods. Our findings demonstrate that both the surface polarization and the polarization plane placement modulate the ion conductance. 

\section{Conclusions}
We study the adsorption, interactions, and current of ions confined in slit-like channels made of polarizable surfaces. The ionic current is non-linear and shows different behavior in channels made of conductors, dielectrics, and non-polarizable surfaces. Strong confinement and polarization effects lead to non-Ohmic ionic currents whereas the two-dimensional transport between conductors tends to be linear. The ionic current modulation and non-linear trend (essential for designing memristors) are caused by the screened or enhanced ionic clustering that is different for each type of confining material. We demonstrate that location of the polarization plane and the confining distance significantly affect the confined fluid properties. Considering the surface of materials with different electronic properties is crucial for studying hybrid ionic/electron coupling in numerous fields such as materials with neuromorphic applications, energy harvesting, and water desalination. In quantum-mechanical calculations, however, considering the coupling between ions transport and the electrons of a surface is challenging due to the differences in time and length scales. Here, we employ the polarization and the polarization plane location to overcome that difficulty in a consistent way with density functional theory.

\begin{acknowledgements}
ACKNOWLEDGEMENTS. A.P.d.S. thanks the Department of Materials Science and Engineering at Northwestern University for appointing him Eshbach Visiting Professor. A.P.d.S. also thanks FAPERGS and CNPq. F.J.-A., A.E., and M.O.d.l.C. acknowledge the support of NSF through the Northwestern University MRSEC grant number DMR-2308691.

A. P. dos Santos and F. Jim\'enez-\'Angeles contributed equally to this work and both are equal first authors.

\end{acknowledgements}

\bibliography{bibliography}

\end{document}